\documentclass[jpcbfk,reprint,superscriptaddress,groupaddress,showkeys]{revtex4-1}

\newcommand{\me}{\mathrm{e}}

\newcommand{\dif}{\mathrm{d}}

\usepackage{graphicx}

\usepackage{amsmath}

\usepackage{amsfonts}

\usepackage{graphicx}

\usepackage{color}

\begin{document}

\title{Molecular-scale Description of SPAN80 Desorption from the Squalane-Water Interface}
\author{L. Tan}
\email{ltan2@tulane.edu}
\affiliation{Department of Chemical and Biomolecular Engineering, Tulane
University, New Orleans, LA 70118}

\author{L. R. Pratt}
\email{lpratt@tulane.edu}
\affiliation{Department of Chemical and Biomolecular Engineering, Tulane
University, New Orleans, LA 70118}

\author{M. I. Chaudhari}
\email{michaud@sandia.gov}
\affiliation{Center for Biological and Material Sciences, Sandia National
Laboratories, Albuquerque, NM 87185}

\date{\today}

\begin{abstract} Extensive all-atom molecular dynamics calculations on the
water-squalane interface for nine different loadings with sorbitan monooleate
(SPAN80), at $T=300$K, are analyzed for the surface tension equation of state,
desorption free energy profiles as they depend on loading, and to evaluate
escape times for absorbed SPAN80 into the bulk phases. These results suggest
that loading only weakly affects accommodation of a SPAN80 molecule by this
squalane-water interface. Specifically, the surface tension equation of state is
simple through the range of high tension to high loading studied, and the
desorption free energy profiles are weakly dependent on loading here. The
perpendicular motion of the centroid of the SPAN80 head-group ring is
well-described by a diffusional model near the minimum of the desorption free
energy profile. Lateral diffusional motion is weakly dependent on loading.
Escape times evaluated on the basis of a diffusional model and the desorption
free energies are $7\times 10^{-2}$~s (into the squalane) and $3\times 10^2$~h
(into the water). The latter value is consistent with irreversible absorption
observed by related experimental work. \end{abstract}

\maketitle

\section{Introduction} 
 
COREXIT 9500 is a standard dispersant used in response to oil spills
\cite{dispersants}. Confronting molecular-scale theory, the formulation of
COREXIT 9500 is non-trivial and has evolved to address issues identified by
decades of experience \cite{riehm2014role}. Therefore, this material provides
context for development of theory, in addition to experiment and simulation,
that might provide molecular insight into the structure and function of
oil-water-surfactant systems.

\begin{figure}\includegraphics[width=2.2in]{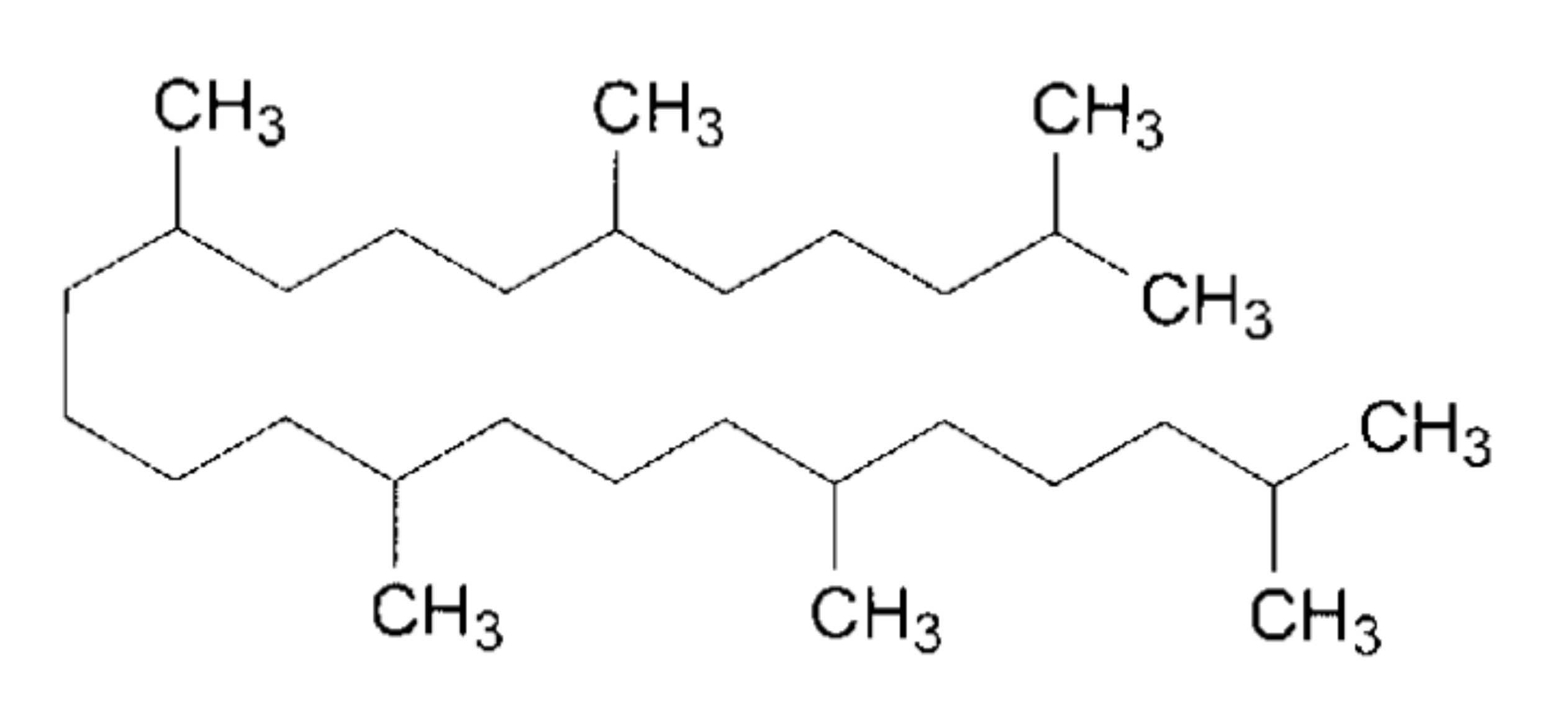}
 \includegraphics[width=3.2in]{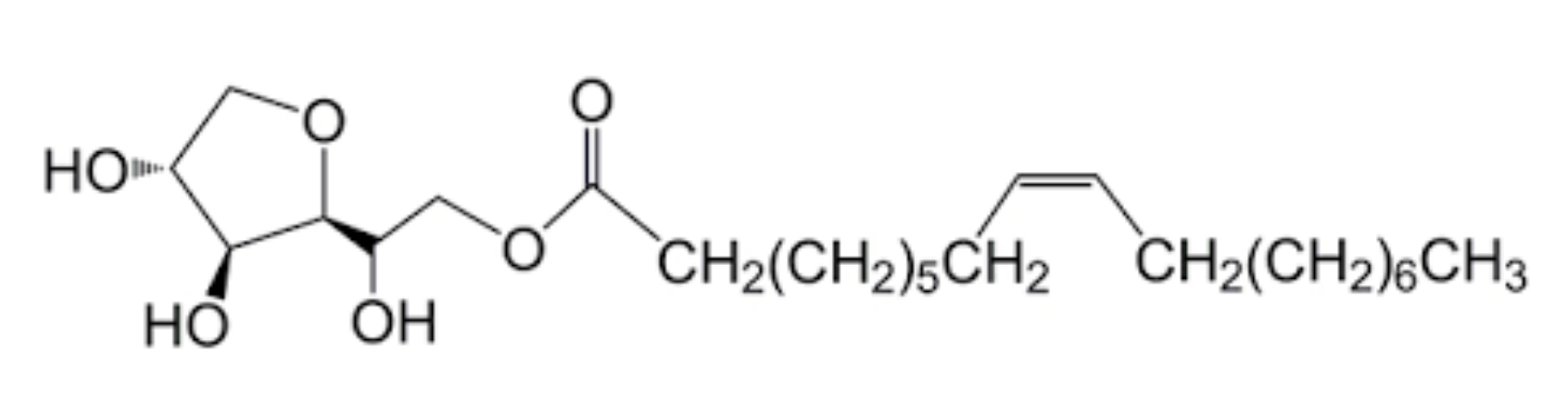}
\caption{Chemical structures of squalane (upper) and sorbitan monooleate (SPAN80, lower).}
\label{fig:SPAN80}
 \end{figure}

COREXIT 9500 includes sorbitan monooleate (SPAN80, FIG.~\ref{fig:SPAN80}), an
ethoxylated sorbitan monooleate (TWEEN80), the anionic surfactant sodium dioctyl
sulfosuccinate (NaDOS), and alkane solvent (NORPAR 13) \cite{dispersants}. Each
of these components, and their proportions, have been chosen to achieve design
characteristics \cite{riehm2014role}. In each case, the statistical mechanical
theory that would provide quantitative molecular-scale explanation is
unavailable. NaDOS provides one example: the molecular theory of electrolyte
effects on the thermodynamics, structure and dynamics of aqueous solution
interfaces is a long-standing
\cite{nichols1982disentanglement,wilson1984hydrophobic,nichols1984salt} and
current research target \cite{jungwirth2014beyond}. As another example, the
specific structural and dynamical description of the ethoxylated chains attached
to the TWEEN80 head-group, and the contrast to the SPAN80 case, is not available
though this is an area of significant recent interest
\cite{Alessi:2005ixa,Norman:2007kq,Norman:2007cg,Chaudhari:2014ga,Chaudhari:2015gt}.

In building the basic molecular theory, it is natural to study the effects
arising from the several components separately, and then to study their various
interactions. Here we study the solution interface between water and a model oil
phase, with SPAN80 at a wide range of loadings. We follow the experimental work
of Reichert and Walker \cite{CMU_Anna} in adopting squalane as the model oil
phase, though they treated TWEEN80 without SPAN80, and water with non-zero NaCl
electrolyte concentration to correspond to seawater. We start with SPAN80 alone
because that is simpler, but further complicating features might be added after
this initial step \cite{kirby2015sequential}.

An intriguing aspect of the experimental work \cite{CMU_Anna} was the
demonstration of irreversible absorption behavior of TWEEN80 for surface
tensions \emph{above} a critical value near 32~mN/m, \emph{i.e.,} for low
surface loadings. That behavior was observed to be independent of the ionic
strength of the aqueous phase. Operational irreversibility might be traced to
rates of desorption, and those rates might be influenced by entanglement of the
ethoxylated head-group of the TWEEN80 surfactant. Evaluating desorption rates
for SPAN80 at this interface provides a baseline result for considering the
TWEEN80 case, and those baseline results are a target of the work which follows
below.

\begin{figure}  \hspace*{-0.5cm} \includegraphics [width=3in]{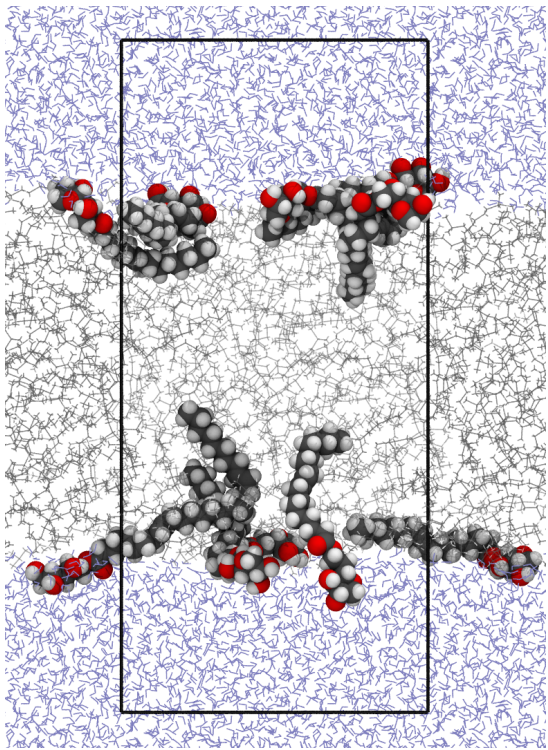}
\caption{System for interface simulations. Here the system included 75
squalane molecules (middle layer), 1000 water molecules (top and bottom), and 5
SPAN80 molecules loaded at each interface. Other cases were composed
similarly but with different numbers of SPAN80 molecules loaded at the
interfaces.} \label{fg:system} \end{figure}

The most important step in determining desorption rates is to establish the free
energy profile for the process. Here, we obtain free energies of desorption of
SPAN80 from the water-squalane interface, stratifying the free energy changes
with a standard \emph{windowing} approach and exploiting parallel tempering
\cite{Earl:2005fv} to investigate sampling sufficiency for the individual
strata. FIG.~\ref{fg:system} shows the case with 5 SPAN80 molecules absorbed at
each interface. The windowing not only provides the free energy profile for the
desorption but also the net free energy for the transfer of a SPAN80 molecule
from water to squalane. Methodological specifics are collected in
Sec.~\ref{simspecifics} below, except where they are pertinent to an isolated
facet of the results. Preceding simulations studied SPAN80 micelles in the
context of drug delivery applications \cite{Han:2013ef}.

A distinguishing aspect of our work is that it is that all-atom models are
treated exclusively, in contrast to coarse-grained models implemented in CHARMM
\cite{CHARMM}, MARTINI \cite{MARTINI} and SDK packages \cite{SDK}. Important
work on escape times of nonionic surfactants from micelles and hydrophobic
surfaces has featured coarse-grained MD models \cite{Ronald}. The present work
is restricted to planar interfaces.

\section{Results and Discussion} 
\begin{figure} \includegraphics[width=3.5in]{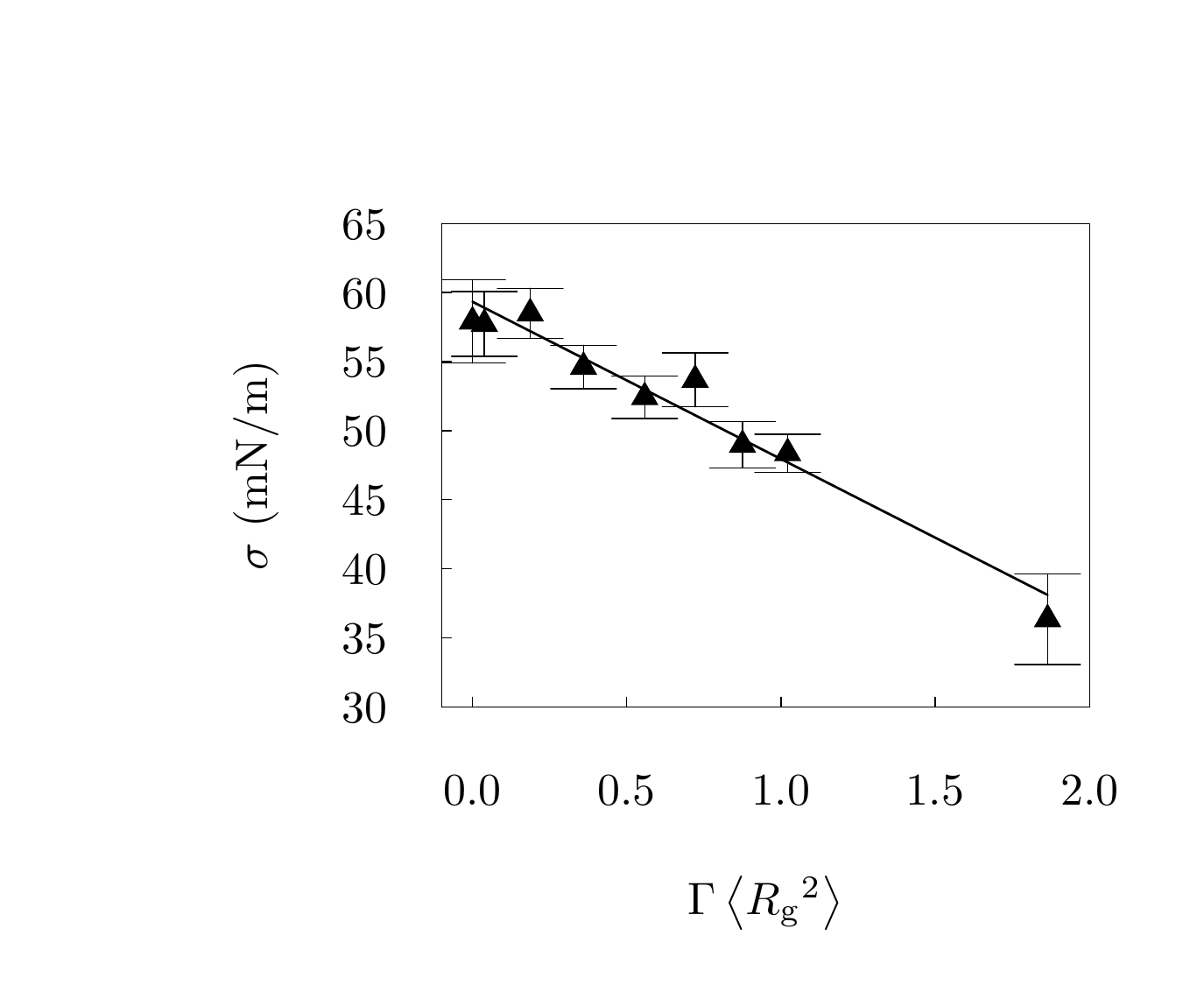} \caption{The surface
tension variation demonstrates the expected trend of decreasing $\sigma$ with
increasing surface desorption $\Gamma$, here scaled by the mean-square radius of
gyration of a SPAN80 molecule in bulk squalane, $ \left \langle R_\mathrm{g}^{2}
\right \rangle ^{1/2}$ = 0.757 nm. These results were produced by standard MD
simulation as described in Sec.~\ref{simspecifics}. $n_{\mathrm{SPAN80}}$ = 0,
1, 5, 10, 15, 20, 25, 30, 50 SPAN80 molecules/interface cases were treated. The
statistical uncertainties indicated are 95\% confidence intervals obtained by a
bootstrap technique based on surface tensions extracted over $\Delta t= 1$~ns
segments along 100~ns simulation for each case. } \label{fg:ST_2} \end{figure}

\begin{figure}
 \begin{center}
 \includegraphics[width=3.5in]{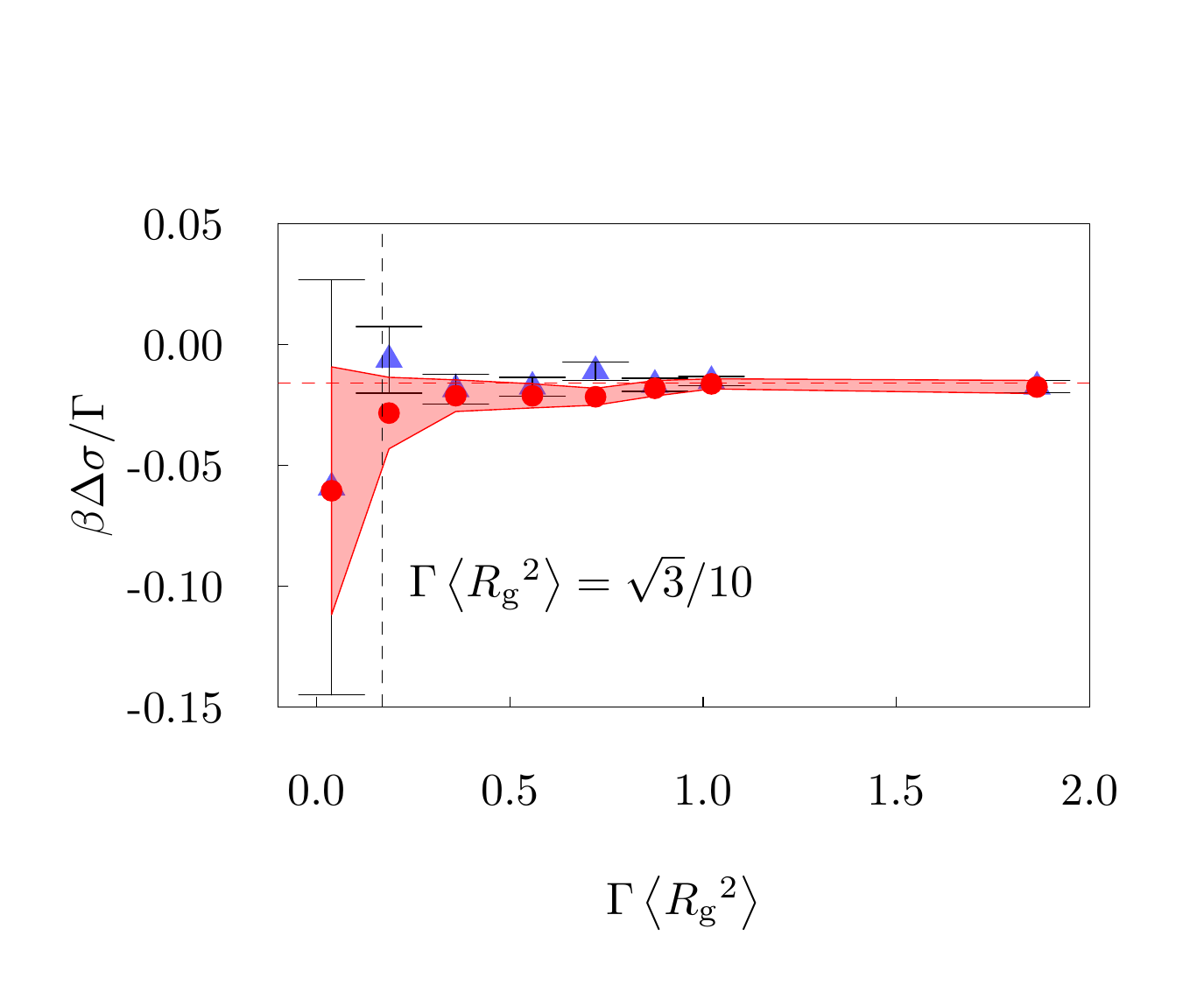} 
 \caption{Dependence of surface stress compressibility factor on surface
loading. The ordinate is the surface tension difference $\Delta \sigma \equiv
\sigma\left(\Gamma\right) - \sigma\left(\Gamma=0\right)$, scaled as indicated
with $\beta = 1/k_\mathrm{B}T$. The red disks are the data from
parallel-tempering simulations as described in Sec.~\ref{simspecifics}. The blue
triangles are the data from standard MD simulations. The vertical dashed line
indicates $\Gamma \left\langle R_\mathrm{g}{}^2\right\rangle = \sqrt{3}/10$ as a
boundary for dilute loading of this interface. This is based on the idea that a
uniform sphere of radius $r$ implies $\left\langle R_\mathrm{g}{}^2\right\rangle
= 3 r^2/5$. If those spheres were closely packed $\Gamma
=1/\left(2\sqrt{3}r^2\right)$. The vertical dashed line marks that value. The
situation of FIG.~\ref{fg:system} corresponds to that low-coverage boundary.} \label{fg:ST_1} 
\end{center}
\end{figure}

Surface tensions were obtained standardly
(Sec.~\ref{simspecifics}) from molecular dynamics simulations for nine surface
loadings (FIG.~\ref{fg:ST_2}). When the absorption $\Gamma$ vanishes, the
surface tension (FIG.~\ref{fg:ST_2}) agrees well with experiment
\cite{CMU_Anna}, and the surface tensions decrease with increasing $\Gamma,$ as
expected. Comparing the standard MD results with parallel tempering values
(FIG.~\ref{fg:ST_1}) shows consistent behavior and remarkable simplicity over
our whole range of loadings. A maximum loading for SPAN80 naturally should be
higher than the maximum loading for TWEEN80, but $\Gamma$ for our strongly
loaded case is about a factor of three (3) higher than the estimated maximum
surface coverage for TWEEN80 \cite{CMU_Anna}. A weakly loaded regime is
identified for $\Gamma < 0.17/\left\langle R_\mathrm{g}{}^2\right\rangle,$ about
10\% of our highest loading.


\begin{figure} \includegraphics[width=3.5in]{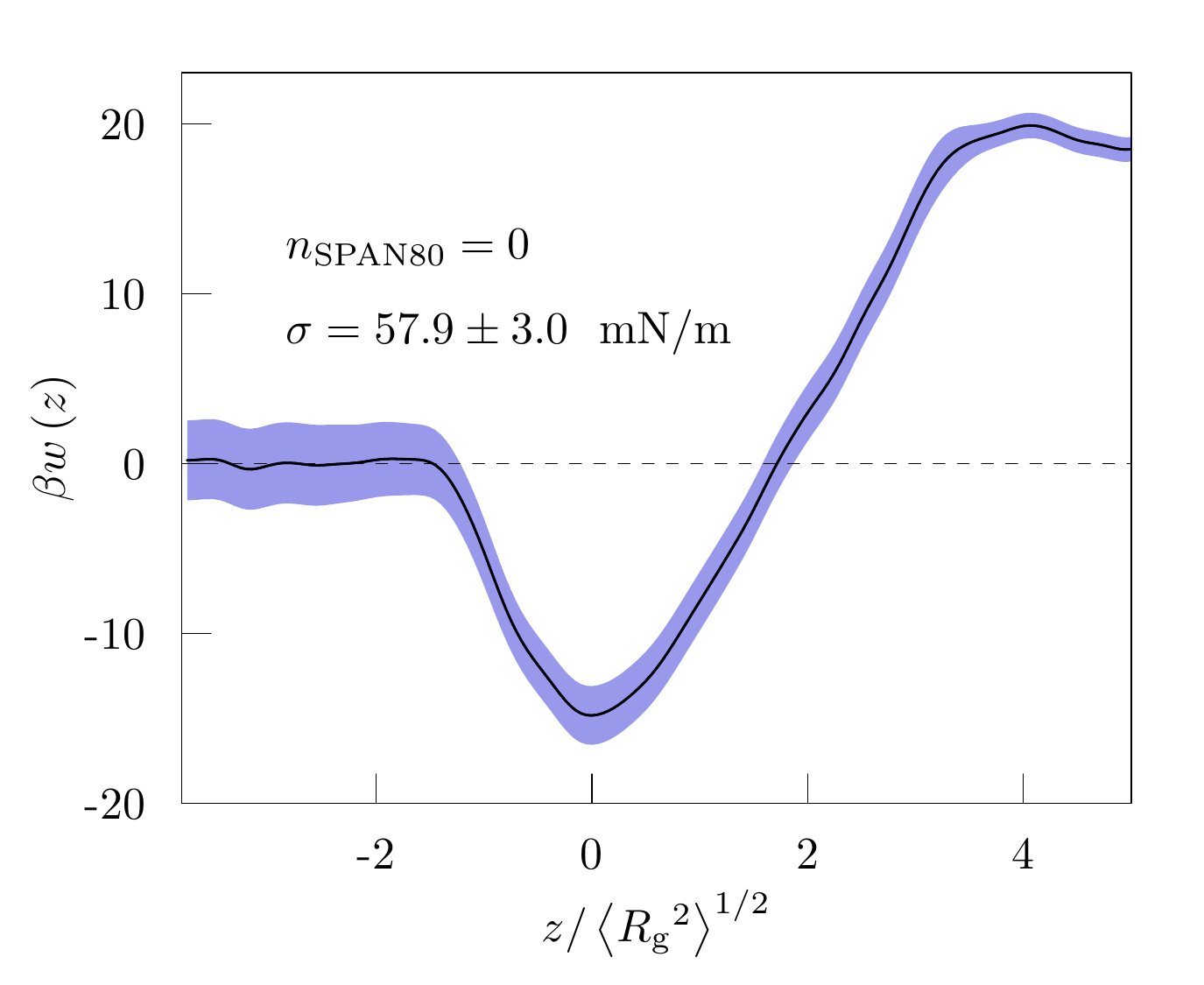} \caption{The
potential of the average force on the centroid of a SPAN80 head-group ring in
the case of the bare water-squalane interface. $ \left \langle R_\mathrm{g}^{2}
\right \rangle ^{1/2}$ = 0.757 nm. The standard stratified sampling approach
used 150 windows used to cover the whole $z$ range (FIG.~\ref{fg:system}).
Calculations for each window ran for 30~ns. $w\left(z\right)$ is reconstructed
by the weighted histogram analysis method. The blue band depicts statistical
uncertainties of $\pm 1$ standard error estimated pointwise on the basis of a
bootstrap resampling of our results. The energy barrier from interface to
squalane phase is around 8.2 kcal/mol and the energy barrier from interface to
bulk water is 19.6 kcal/mol.} \label{fg:pmf_Emp} \end{figure}

\begin{figure}
\includegraphics[width=3.5in]{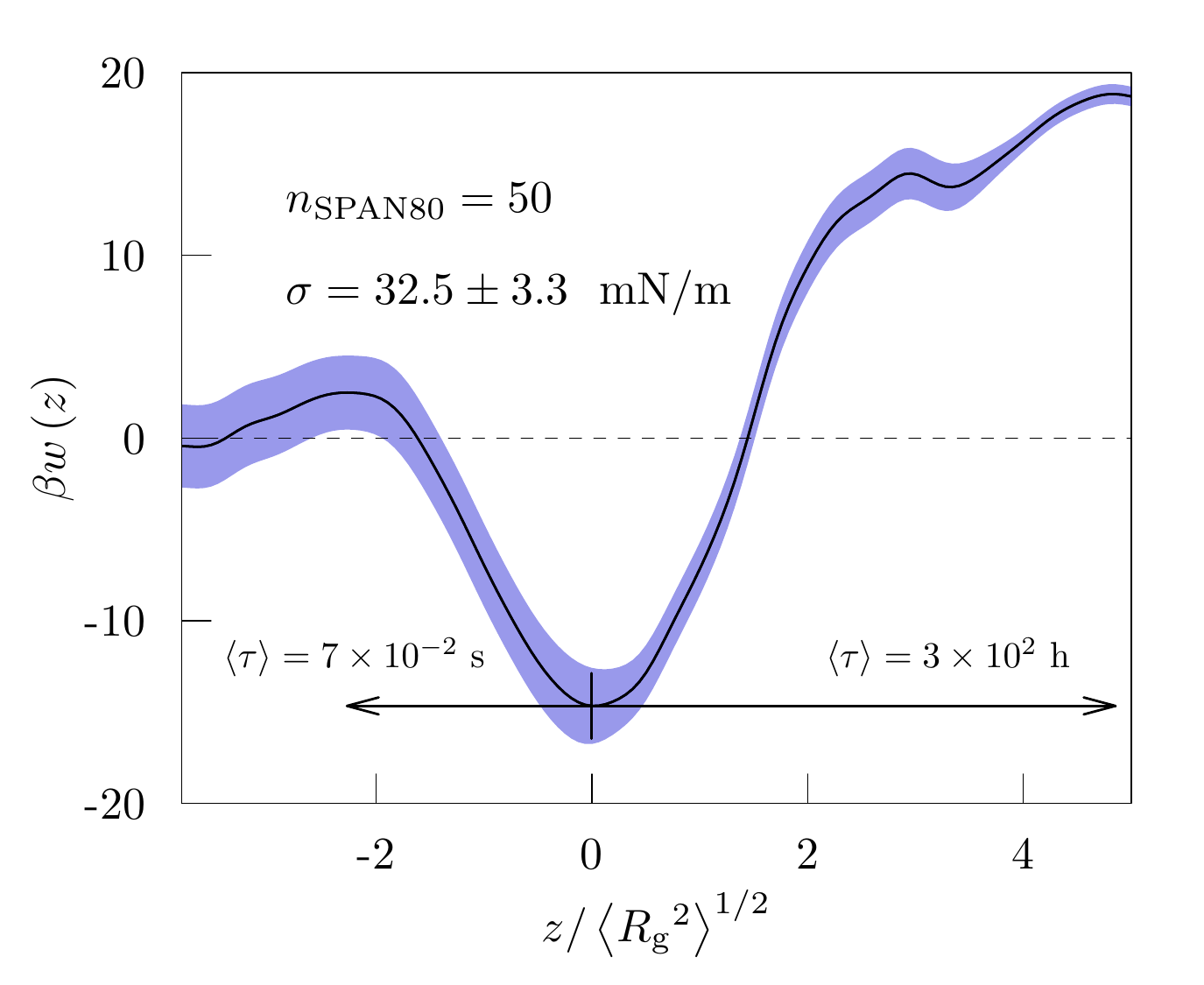}
\caption{The potential of mean force for the strongly loaded water-squalane
interface. The abscissa is the $z$ coordinate of the centroid of a SPAN80
molecule head-group ring scaled by mean-square radius of gyration of a SPAN80
molecule in bulk squalane, $ \left \langle R_\mathrm{g}^{2} \right \rangle
^{1/2}$ = 0.757 nm. The standard stratified sampling approach used 190
windows to cover the whole $z$ range. Calculations for each window ran for
60~ns. $w(z)$ is recomposed by the weighted histogram analysis method. The blue
band depicts statistical uncertainties of $\pm 1$ standard error estimated
pointwise on the basis of a bootstrap resampling of our results. These results
show an energy barrier from interface to squalane phase is around 8.7~kcal/mol
and the energy barrier from interface to bulk water is around 19.6~kcal/mol.}
\label{fg:pmf_Full}
\end{figure}

Desorption free energy profiles were obtained for the unloaded case
($n_{\mathrm{SPAN80}}$ = 0/interface) and our strongly loaded case
($n_{\mathrm{SPAN80}}$ = 50/interface).  The free energy profiles
(FIGs.~\ref{fg:pmf_Emp} and \ref{fg:pmf_Full}) are remarkably similar for the
two cases. The desorption free energy barrier from interface to squalane phase
is around 8 kcal/mol and the free energy barrier is 19.6 kcal/mol from interface
to bulk water.   

To analyze the kinetics of the desorption process, we assume
that the motion of the centroid of a SPAN80 molecule head-group ring may be
described by the Smoluchowski equation 
\begin{eqnarray} \frac{\partial}{\partial
t} P(z,t \vert z_{0}) = D \frac{\partial}{\partial z} \left( \beta
w^\prime\!\left(z\right) +\frac{\partial}{\partial z}\right)P(z,t \vert z_{0})
~, \label{eq:fokker} \end{eqnarray} 
with $P(z,t \vert z_{0})$ the conditional probably density for a centroid of a
SPAN80 molecule head-group ring to arrive at $z$ after a time $t$ from an
initial location $z_0$ \cite{ngvK}. Motions 
parallel to the interface, in the $x$ and $y$ directions,  are separable in this description. We will use this
basis to evaluate the mean first passage time for escape of a SPAN80 molecule from
the interface. This description uses $w\left(z\right)$ obtained above, but
requires also the kinetic parameter $D$, a self-diffusion coefficient.

\begin{figure} \hspace*{-1cm}\includegraphics[width=3.5in]{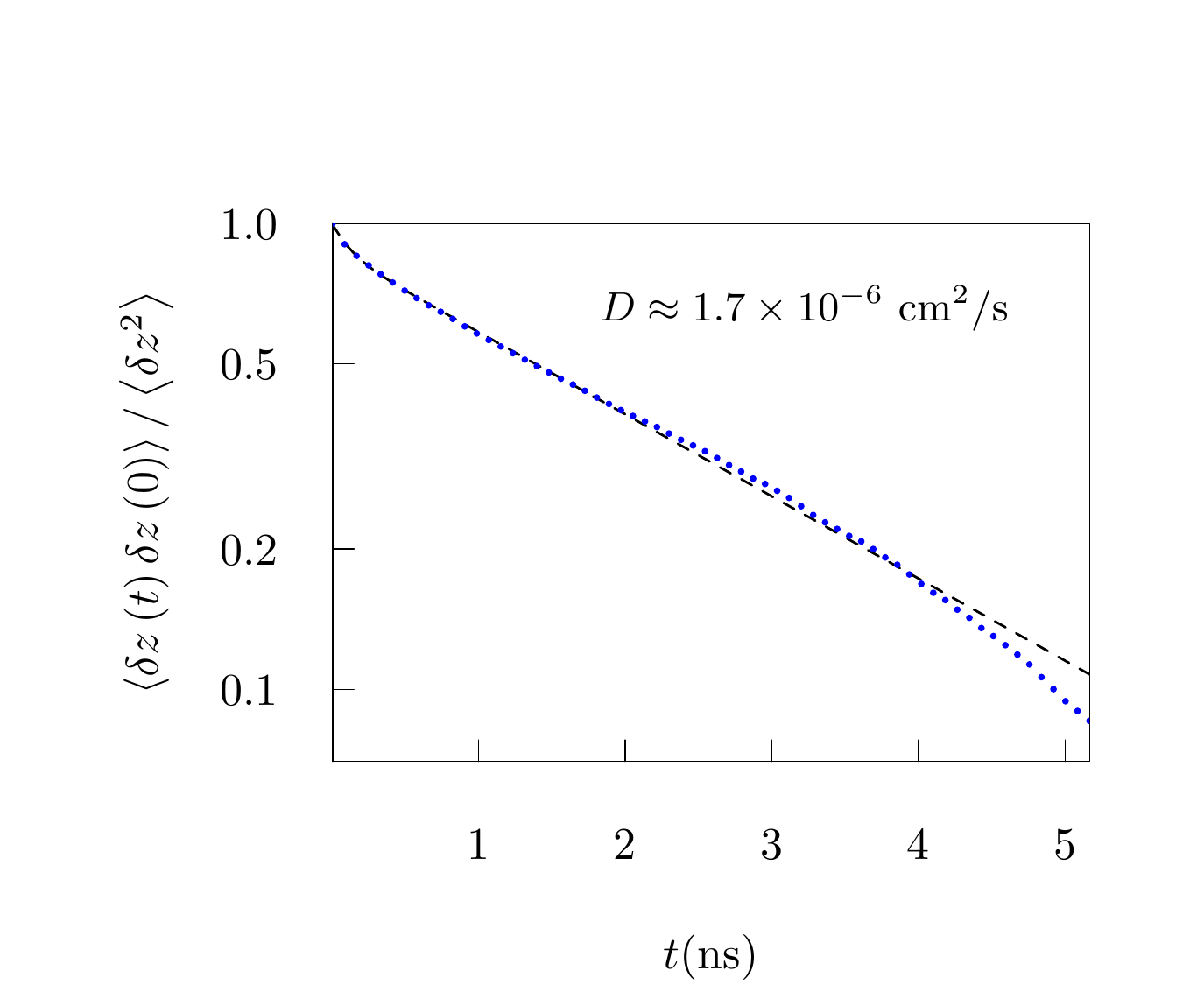}
\caption{The autocorrelation function for displacement of the centroid of a
SPAN80 head-group ring at the strongly loaded interface. The black dashed line
is the function $ A\me^{-t/\tau_1} + (1 - A)\me^{-t/\tau_2}$ fit to the
primitive results (blue dots). $A = 0.12,$ $\tau_1=0.12$~ns, and
$\tau_2=2.46$~ns.} \label{fg:diffusion} \end{figure}

To evaluate $D$, we ran another standard MD simulation preserving sufficient time
resolution. Specifically, we extracted a windowed configuration from the
strongly loaded interface simulation, then extended the MD simulation for 100~ns
with configurations saved every 10~fs. The first 20~ns of this trajectory was
discarded as further aging, and we naturally used results for all SPAN80
molecules present. The centroid of a SPAN80 head-group ring is followed with
$\delta z=0$ locating the minimum of $w(z)$ (FIG.~\ref{fg:pmf_Full}). The
centroid of a SPAN80 molecule head-group ring wanders near the minimum of
$w\left(z\right)$, and we linearize $w^\prime\!\left(z\right) \approx \kappa
\delta z.$ The Langevin equation 
\begin{eqnarray}
\delta \dot{z}(t) + \beta \kappa D \delta z(t) = R(t)~, \label{eq:Smoluchowski}
\end{eqnarray} 
with $R(t)$ the random-force exhaustively discussed elsewhere \cite{fpt}, then
corresponds to Eq.~\eqref{eq:fokker}. Establishing the necessary force constant
by $\beta\kappa \left\langle \delta z^2\right\rangle =1$ gives 
\begin{eqnarray}
\left\langle \delta z(0)\delta z(t) \right\rangle =  \left\langle \delta z^2 \right\rangle\me^{-D t/\left\langle
\delta z^2 \right\rangle}~.
\label{eq:average3} \end{eqnarray}
 The observed displacement time-correlation-function (FIG.~\ref{fg:diffusion})
relaxes exponentially, confirming the basic kinetic description. The slight
deviation from exponential relaxation (FIG.~\ref{fg:diffusion}) could be due to
the linearization $w^\prime\!\left(z\right) \approx \kappa \delta z.$ The
self-diffusion coefficient is $1.7 \times 10^{-6} \mathrm{cm^{2}/s}$, about a
factor of 10 less than a corresponding value for liquid water, and about half of
the self-diffusion coefficient of propylene carbonate \cite{you2014role}.
Experimental results for diffusion coefficients of nonionic surfactants range
from $1\times 10^{-6} \mathrm{cm^{2}/s}$ to $8\times 10^{-6} \mathrm{cm^{2}/s}$
\cite{Tween20, Miller, Hasko, Moorkanikkara}. The present value corresponds to
relaxation times $\left\langle \delta z^2 \right\rangle/D$ of about 2~ns
(FIG.~\ref{fg:diffusion}). Lateral diffusion is slightly faster here
(FIG.\ref{fg:xydif}), but slows slightly with increasing interface loading.

With this value of $D$, we adapt the theory of first passage times \cite{fpt} to 
\begin{eqnarray}
\left\langle \tau \right\rangle =  \left(\frac{2 \pi\left\langle \delta z^2 \right\rangle }{D}\right)
 \int_{0}^{\delta z^\ddag}\me^{\beta\Delta w(z)}
 \frac{\dif \delta z }{\sqrt{2\pi \left\langle \delta z^2 \right\rangle}}
\label{eq:FPT}
\end{eqnarray}
where $\Delta w(z) = w(z) - w_{\mathrm{min}}$, and $\delta z^\ddag$ is the
displacement from the minimum at $\delta z=0$ to the barrier configuration. We
then find (FIG.~\ref{fg:pmf_Full}) the mean first passage time of 0.07~s, from
the interface to the squalane phase, while the mean first passage time from the
interface to bulk water is about $3\times 10^{2}$~h. Differences in the barrier
shapes in the two directions play a role in determining that value, since the
ratio of these times is about a factor of four less than the naive ratio of
$\me^{\beta \Delta w^\ddag} \approx \me^{18}.$

\section{Conclusions}  

\begin{figure}
\includegraphics[width=3.5in]{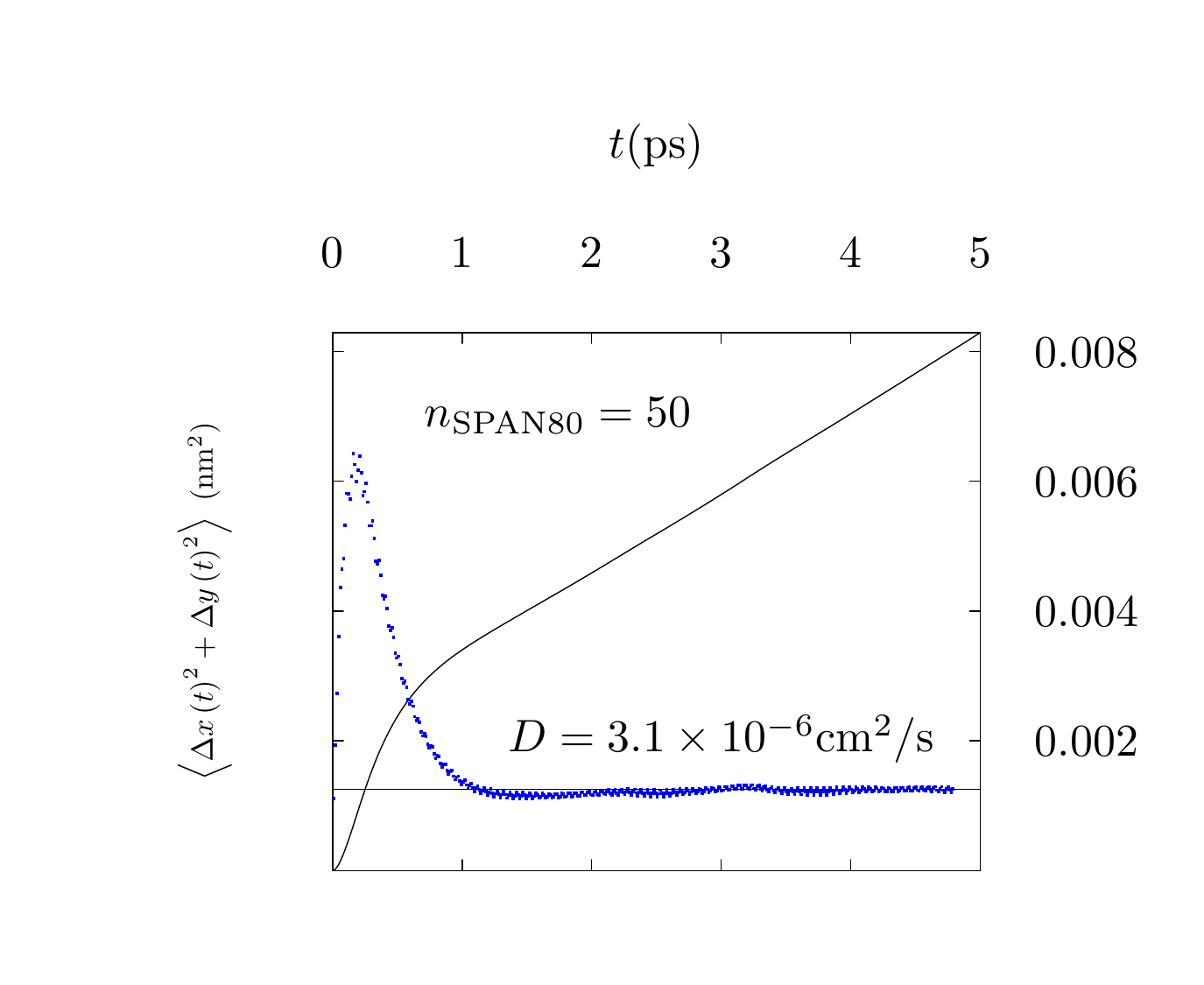}
\includegraphics[width=3.5in]{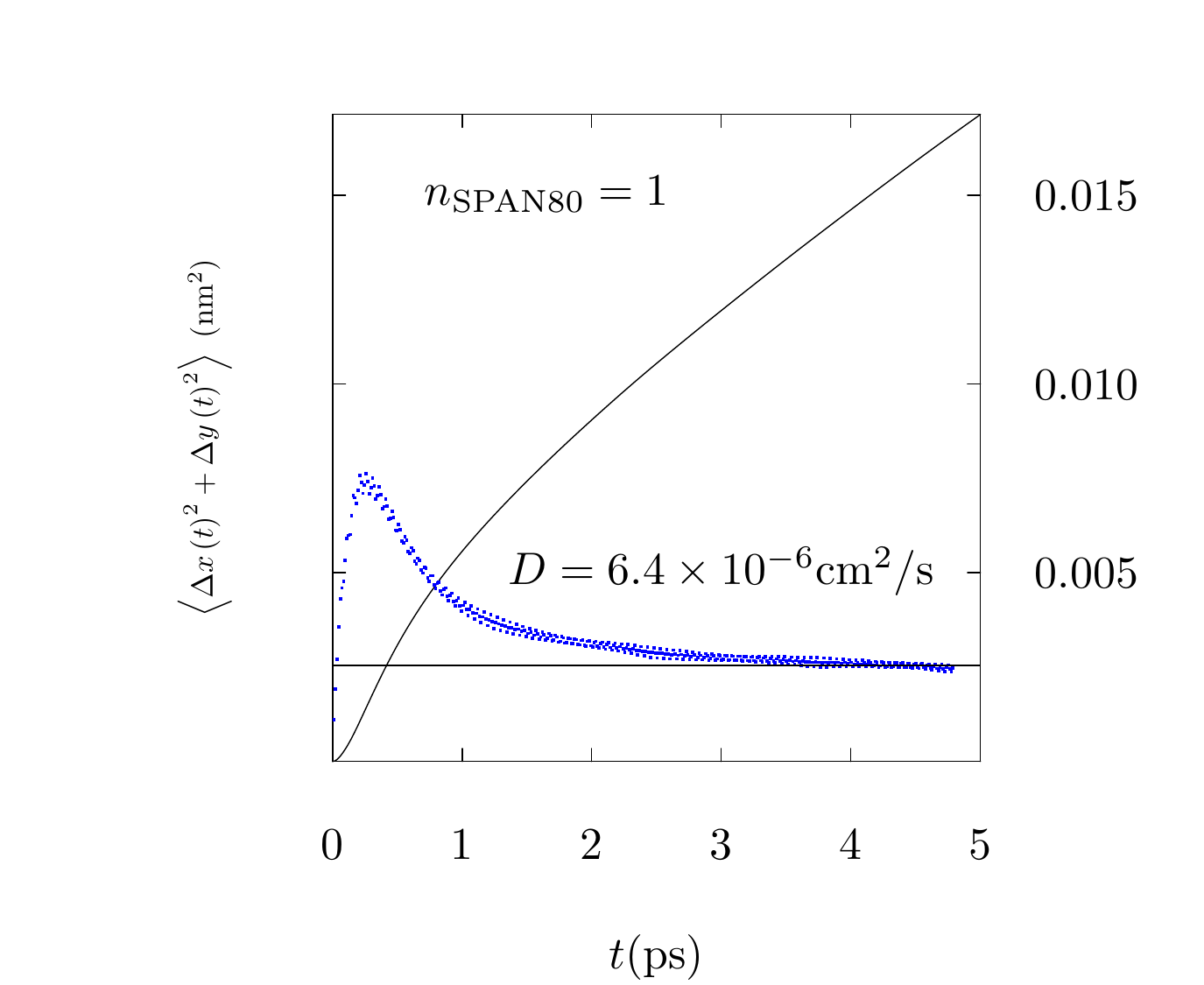}
\caption{Lateral diffusive motion of the SPAN80 ring centroid and its dependence
on loading. The plotted blue marks are estimates of the time derivative of the
solid lines. Lateral diffusion coefficients are slightly larger than the
perpendicular coefficient (FIG.~\ref{fg:diffusion}), but interfacial crowding
has the expected effect of slowing the diffusion slightly.}
 \label{fg:xydif} \end{figure}

These results suggest that loading only weakly affects accommodation of a SPAN80
molecule by this squalane-water interface. Specifically, the surface tension
equation of state (FIG.~\ref{fg:ST_1}) is simple through the range of high
tension to high loading studied, and the desorption free energy profiles are
weakly dependent on loading here (FIGs.~\ref{fg:pmf_Emp} and \ref{fg:pmf_Full}).
The free energy of transfer of a SPAN80 molecule from water to squalane is about
-28~kcal/mol. The perpendicular motion of the centroid of the SPAN80 head-group
ring is well-described by a diffusional model near the minimum of the desorption
free energy profile (FIG.~\ref{fg:diffusion}). Lateral diffusional motion is
weakly dependent on loading (FIG.~\ref{fg:xydif}). Escape times approximated on
the basis of a diffusional model and the desorption free energies are $7\times
10^{-2}$~s (into the squalane) and $3\times 10^2$~h (into the water), the latter
value consistent with irreversible absorption observed by recent experimental
work \cite{CMU_Anna} on a related system.

\section{Methodological Specifics}\label{simspecifics} The GROMACS \cite{gromacs}
4.6.7 molecular dynamic simulation package was used for all calculations. The
chain molecules were represented by OPLS-AA force field \cite{oplsaa} and the
SPC/E model \cite{spce} was chosen for water. The Nose-Hoover thermostat
\cite{nosehoover} maintained the constant temperature and Parrinello-Rahman
barostat \cite{rahman} kept pressure at 1.0~atm. Long-range electrostatic
interactions were treated in standard periodic boundary conditions using the
particle mesh Ewald method with a cutoff of 1.0~nm. The chemical bonds involving
hydrogen atoms were constrained by the LINCS algorithm \cite{lincs}. 

In evaluating surface tensions, we applied standard MD simulation and checked
the sampling sufficiency with parallel tempering calculations. Nine different
surface loadings were investigated, with $n_{\mathrm{SPAN80}} = 1, 5, 10, 15, 20
,25, 30$ per interface. The squalane phase included 75 squalane molecules and
was bounded by a water phase, bottom and top layers, of 1000 water molecules.
For the highest loaded ($n_{\mathrm{SPAN80}} = 50$/interface) and unloaded
($n_{\mathrm{SPAN80}} = 0$/interface) cases, a larger system of 100 squalane
molecules and 3000 water molecules was simulated. We expanded the system for
those two cases to accommodate a consistent comparison for desorption free energy
evaluations. Our standard procedure carried-out an energy minimization
calculation and density equilibration followed by a 100~ns production run with
constant particle number, pressure, and temperature (NPT) conditions. The
parallel tempering calculations used 48 replicas spanning the 260-450K
temperature range. Trajectories ran for 50~ns and the resulting exchange
probability between neighboring temperatures was around 20\%. 

We utilized the windows sampling approach to evaluate the desorption free energy
profile. To generate initial configurations for each window, we placed one more
SPAN80 in the water phase at the position where $z =1.0$~nm. A pulling force was
applied to the centroid of the SPAN80 head-group ring to pull it across the
whole system. To achieve high resolution for $w(z)$, 150 windows were used to
cover the whole $z$ range in unloaded system while 190 windows are utilized to
cover the whole $z$ range in strongly loaded system. Due to the different
complexity between two cases, the extent of the MD trajectory/window differs.
Trajectories ran for 30~ns/window in unloaded case while the calculations
extended to 60~ns/window in the strongly loaded case.

\section{Acknowledgement} Sandia is a multiprogram laboratory operated by Sandia
Corporation, a Lockheed Martin Company, for the U.S. Department of Energy's
National Nuclear Security Administration under Contract No. DE-AC04-94AL8500.
The financial support of Sandia's LDRD program and the Gulf of Mexico Research
Initiative (Consortium for Ocean Leadership Grant SA 12-05/GoMRI-002) is
gratefully acknowledged.



\providecommand{\latin}[1]{#1}
\providecommand*\mcitethebibliography{\thebibliography}
\csname @ifundefined\endcsname{endmcitethebibliography}
  {\let\endmcitethebibliography\endthebibliography}{}

\end{document}